**Prime novelty**

Polycrystalline diamond coatings with uniform structure and thickness distribution are grown by microwave plasma CVD on WC-6%Co substrates using an optimized pocket-type substrate holder. Wear rate of the coated cutting inserts upon cutting of Si-Al abrasive alloy was strongly reduced.

**Highlights**

● Multiple growth of diamond coatings on WC-Co substrates by MPCVD is realized

● Coatings with uniform thickness are produced using a pocket-type substrate holder

● Diamond grain size and roughness uniformity are achieved

● Diamond-coated WC-6%Co cutting inserts are tested upon cutting of A390 Si-Al alloy

● Cutting force and wear rate are strongly reduced for the diamond-coated tool

**Uniform diamond coatings on WC-Co hard alloy cutting inserts deposited by a microwave plasma CVD**

E.E. Ashkihazi[1,2], V.S. Sedov[1,2], D.N. Sovyk[1,2], A.A. Khomich[1,3], A.P. Bolshakov[1,2], S.G. Ryzhkov[1], A.V. Khomich[1,3], D.V. Vinogradov[1,5], V.G. Ralchenko[1,2,4], V.I. Konov[1,2].

[1]General Physics Institute of Russian Academy of Sciences, Vavilov Str. 38, 119991 Moscow, Russia

[2]National Research Nuclear University "MEPhI", Kashirskoe Shosse 31, 115409 Moscow, Russia

[3]Institute of Radio Engineering and Electronics RAS, 141190 Fryazino, Russia

[4]Harbin Institute of Technology, 92 Xidazhi Str., 150001 Harbin, P.R. China

[5] Bauman Moscow State Technical University, 2nd Bauman Str.5, 105005 Moscow, Russia

**ABSTRACT**

Polycrystalline diamond coatings have been grown on cemented carbide WC-6%Co substrates with different aspect ratios by a microwave plasma CVD in $CH_4/H_2$ gas mixtures. To protect the edges of the substrates from non-uniform heating due to the plasma edge effect, a special plateholder with pockets for group growth has been used. The difference in heights of the substrates and plateholder, and its influence on the diamond film mean grain size, growth rate, phase composition and stress was investigated. The substrate temperature range, within which uniform diamond films are produced with good adhesion, is determined. The diamond-coated cutting inserts produced at optimized process exhibited a reduction of cutting force and wear resistance by a factor of two, and cutting efficiency increase by 4.3 times upon turning A390 Al-Si alloy as compared to performance of uncoated tools.

**Keywords:** diamond film, microwave plasma CVD, cemented carbide, cutting tools, wear resistance

**1. Introduction**

Diamond coatings on hard alloy tungsten carbide tools, owing to low friction coefficient, high strength and wear resistance, are of great interest for treatment of advanced materials [1, 2]. Most commonly the diamond coatings on cutting tools are produced by chemical vapor deposition (CVD) using a microwave plasma (MPCVD) and hot filament (HFCVD) [3]. While the MPCVD technique principally provides a higher growth rate and film quality compared to HFCVD, the process stability and the film uniformity strongly depend on the shape and aspect ratio of the substrates. Tungsten needles-cathodes [4] for field electron emission, drills and inserts for superhard tools [5] and others are the examples of high-aspect ratio (h/d, where h means thickness and d means diameter) substrates. The growth in microwave (MW) plasma on substrates with high enough aspect ratio (h/d > 0.1) requires special means to provide a uniform heating [4, 6]. The MW plasma shape is sensitive to substrate geometry because of an edge effect: the plasma concentrates in the areas of electric field distortions, especially at the edge of strongly protruding substrate that causes a non-uniform diamond film deposition [6]. This detrimental effect of local MW field and the substrate temperature disturbance is aggravated with aspect ratio increase [7, 8], for example, in case of drills. Therefore, the temperature uniformity becomes a key condition determining the possibility of successful diamond deposition. Overheating of edges of a WC-Co substrate considerably facilitates cobalt diffusion which, penetrating the surface catalyses non-diamond ($sp^2$) phase formation, destroying the adhesion [2] between WC-Co and diamond film. The temperature rise degrades the adhesion because it increases an internal stress at the boundary of diamond film and substrate due to the difference in coefficients of thermal expansion.

The transition from the use of a microwave reactor with a frequency of 2.45 GHz to reactors with 915 MHz frequency improves heating uniformity [9], but only for low-aspect substrates (h/d < 0.1). One can avoid this problem using HFCVD and DC electrical discharge systems with more uniform temperature distribution [3, 10-12] to produce diamond films on complex shape substrates. However, there are technical difficulties because of filament carburization [13], moreover, typical growth rates in the HFCVD method are by an order of magnitude lower compared to that for MPCVD [14]. The heating uniformity becomes to be a problem upon MPCVD deposition scaling. So, for a single substrate as well as for group deposition the temperature of a substrate depends on its height and location on the substrate holder [15, 16]. It was suggested to lower the temperature gradient for single high-aspect ratio substrates h/d = 0.25 [17] and h/d = 0.4 [6]) owing to use special substrate holders.

Here, we obtained a uniform heating of high aspect ratio WC-6%Co hard alloy substrates by eliminating the edge effect using a cassette-type substrate holder (plateholder). Diamond films growth on substrates of different aspect-ratios is investigated, and the morphology, grain size, growth rate and internal stress of diamond films are analyzed. Finally, the diamond-coated cutting inserts produced at optimized conditions were tested upon turning A390 Al-Si alloy.

## 2. Experimental

Three series of experiments on diamond deposition were performed using WC-6%Co substrates of cylindrical shape with diameter d = 10 mm and different heights h of 2.5, 3.5, 4.5 and 5.5 mm (the aspect ratio h/d ranged from 0.25 to 0.55). The samples were placed on the substrate holder around its symmetry axis (Fig. 1). Preliminary, the substrate surface was polished, then chemically etched and coated with a tungsten diffusion barrier layer. Cobalt content in subsurface layer was reduced by selective two-stage etching with Murakami mixture ($K_3[Fe(CN)_6]$: KOH : $H_2O$ = 1 : 1 : 10) and Caro mixture (3 ml 96% w/w $H_2SO_4$ : 95 ml 37% w/w $H_2O_2$ in water). This etching procedure resulted in a well-developed relief that enhanced adhesion of the thin tungsten barrier layer. The tungsten interlayer with thickness of 0.6 µm was deposited by magnetron sputtering at DC current of 0.5 A and voltage of 360 V in argon (99.998%) atmosphere at pressure of 0.5 Pa and WC-Co substrate temperature of 50$^o$C. We used home-made facility to maintain these deposition conditions. Before the magnetron sputtering the argon ion beam cleaning was performed at current density of 1.0 mA/cm$^2$ and energy of 500 eV. We use ion source which is develop and produce by ourselves. These ion guns belong to the gridless type – the so-called anode layer ion source. The first series of experiments was performed for a group of four samples in an open-type substrate holder made of molybdenum (Fig. 1 a,b,f), in conditions, when the plasma embraced all the surface of the substrates. The second and third series were performed in a closed-type plateholder with pocket placed on the flat Mo holder (Fig. 1 c,d,e,g,h). The height of all the substrates was the same $h_0 = 4.5$ mm in the first and the third experiments (Fig. 1 f,h), and it was reduced by 1 or 2 mm to $h_1 = 3.5$ mm and $h_2 = 2.5$ mm, or increased to $h_4 = 5.5$ mm in the 2$^{nd}$ experiment (Fig. 1 c). Morphology and phase composition of samples with $h_0 = 4.5$ mm from the 2$^{nd}$ and 3$^{d}$ series it turned out to be similar. We chose one of them from the 3$^{d}$ series for cutting test. The plateholder height was $H_{p-h} = 3.5$ mm. The samples and, therefore plateholder height were chosen as high as mean height of commercial WC-Co cutting inserts. Thus, there were the following variants of the samples height relative to the pocket-type holder: $\Delta h_1 = -1$ mm, $\Delta h_2 = 0$ mm, $\Delta h_3 = 1$ mm, $\Delta h_4 = 2$ mm (Table 1).

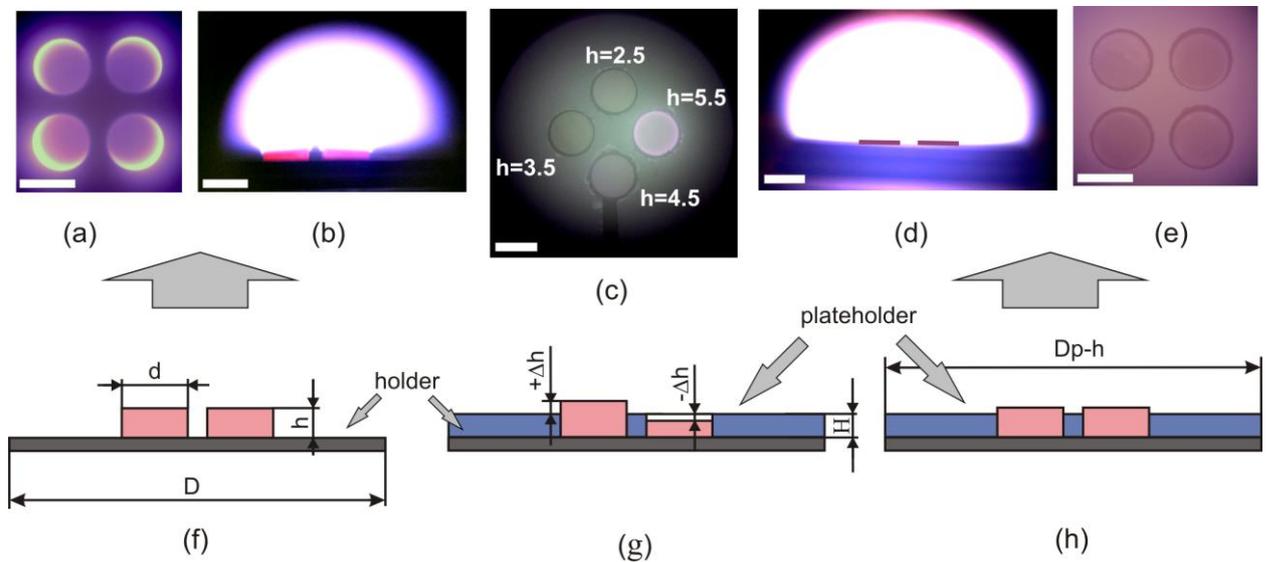

Fig. 1. Photographs of the samples in open substrate holder during growth, top view (a) and side view (b) – first series of diamond CVD; a photograph of substrates of different heights in the plateholder pockets during diamond growth, top view (c) – second series of diamond CVD; a photograph of substrates with equal heights in the plateholder during growth, side view (d) and top view (e) – third series of diamond CVD; schematic views of the substrates arrangement in the open holder, h/d = 0.45 (f) – first CVD series; in the close-type plateholder with substrates of different height (g) – second CVD series and equal height h/d = 0.45 (h) – third CVD series. The plateholder diameter is $D_{p-h} = 57$ mm and height is H = 3.5 mm. White scale bars size is 10 mm in the lower left corner at figures (a)-(e).

We grew diamond films in home-made "Ardis-100" MPCVD reactor chamber [14] similar to "Astex" (now "SEKI-Diamond") system. Magnetron frequency was 2.45 GHz, but microwave energy delivers coaxially to the substrate from the bottom. The diamond deposition conditions in all the experiments were as follows: methane concentration in $H_2$ of 4%, microwave power of 2.9 kW, pressure of 70-80 Torr, and the deposition time of 4 hours. We measured the substrate temperature both from the top, through the plasma, by a Mikron M770 pyrometer with space resolution of some mm, and from the lateral side (Fig. 1c), through a slot in the plateholder wall under weakened plasma discharge illumination, by a Willamson 81-35-C two-colour pyrometer. The film thickness was calculated by the mass increment due to the film growth. Scanning electronic microscopy investigations were carried out by SEM JEOL 7001F, the elemental analysis by X-ray fluorescence (XRF) spectrometer INCA Energy X-MAX, the analysis of phase composition, by a LabRAM HR 800 micro-Raman spectrometer with excitation at 473 nm and space resolution of 1 μm, the optical profilometry (roughness measurements), by a Zygo NewView 5000 instrument, and the cutting force and wear rate when machining A390 silumin were measured by an instrument based on an automatic screw machine 1K62 equipped with a Kistler 9257B dynamometer. A silumin workpiece of 85 mm in diameter was machined under the following conditions: a cutting speed of $V = 220$ m/min, a longitudinal feed of $S = 0.15$ mm/r, and a cutting depth of $t = 0.5$ mm. The limit wear of the cutter was 0.5 mm in all cases. Contact profilometer Calibre 252 with diamond tip was applied to measure curvature radius of the cutting edges.

## 3. Results and discussion

When one uses an open plateholder, the heating of the substrate is strongly non-uniform (Figs. 1 a,b): the outer edge of the samples is overheated due to the increased microwave power density (edge effect); as a result, the diamond film peels off in the overheated areas. It was noticed that microwave plasma discharge is formed on each high-aspect substrate and is modified depending its position in the reactor cavity (Fig. 1b). This leads to non-uniform heating of samples even in a simple case with four symmetrically placed substrates of identical height. To eliminate this negative effect, we changed the shape of microwave plasma. Instead of a contour formed on individual substrates, we produced a common contour of plasma cloud with base on the perimeter of the plateholder (Figs. 1 c-h). The diameter D of the plateholder was chosen equal to half of the wavelength λ at microwave frequency of 2.45 GHz, D = 57 mm ≈ λ/2, and the height of the plateholder was chosen equal to the average height of the experimental samples of substrates (H = 3.5 mm). The uniformity of temperature at the center and edges of samples on the growth plane was qualitatively estimated by the brightness of the colour range of the photo, which was the same on all the substrates in the plateholder (Fig. 1e). The geometry of substrates of different height and the parameters of microcrystalline diamond films deposited at different temperatures in microwave plasma are shown in Table 1. The parameters highlighted in the table guarantee deposition conditions with good adhesion and uniform morphology of diamond films on WC-6% Co substrates.

Table 1. Deposition conditions and properties of the diamond coatings (data for the samples of h = 2.5 mm, 3.5 mm, 5.5 mm: second CVD series; data for the sample of h = 4.5 mm: third CVD series).

| Substrate height h, mm | 2.5 | 3.5 | 4.5 | 5.5 |
|---|---|---|---|---|
| Plateholder height H(p-h), mm | 3.5 | 3.5 | 3.5 | 3.5 |
| Δh, mm | -1 | 0 | 1 | 2 |
| Temperature from the side T(side), °C | 700 | 740 | 760 | 870 |
| Temperature from the top T(top), °C | 780 | 800 | 825 | 960 |
| Thickness of diamond film, μm | 0.6 | 1.2 | 4.1 | 13.8 |
| Grain size d(grain), μm | 0.40 | 0.55 | 0.80 | 2.2 |
| Growth rate, μm | 0.15 | 0.3 | 1.0 | 3.4 |
| Compressive stress σ, GPa | 0.68 | 2.2 | 2.5 | 3.0 |

Traditionally, the substrate temperature T(top) measured from the top through plasma is systematically overestimated due to the radiation of microwave plasma, which is a gray body [18]; therefore, more correct values of temperature are those measured from the lateral side, T(side). The temperature difference T(top) – T(side) reached 60-90°C in our case at pressures (70-80 Torr). It turned out that the experimental temperature of substrates (Табл. 1) is very sensitive to the variation in the parameter Δh (Fig. 2).

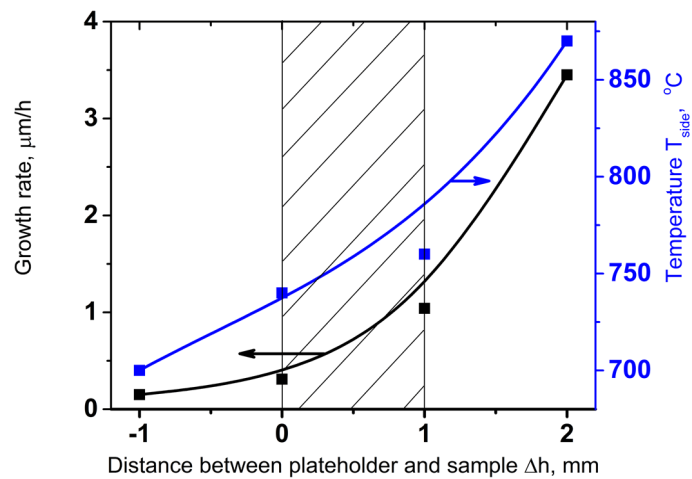

Fig. 2. The growth rate and temperature T(side) of diamond films deposited on WC-6% Co substrates in microwave plasma as a function of the distance Δh between the substrate and the plateholder surfaces. The optimum CVD region is shown by hatching.

The temperature interval and the range of distances within which the diamond film deposited on WC-6% Co grows at a rate of 0.3-1.0 µm/h and does not peel off appear to be ΔT(side)=20 °C (ΔT(top)=25 °C) and Δh = 0…+1 mm (Fig. 3). At low temperatures, the film grows too slowly; for example at 700°C (Δh = – 1 mm) the growth rate is 0.15 µm/h (Table 1), and it does not grow continuous even for 4 hours (Fig. 4a). On the contrary, at higher temperature T(side) = 870°C, when the substrate height exceeds the plateholder height by Δh = +2 mm, the film grows quite rapidly (3.4 µm/h, Table 1 and Fig. 4d), but it peels off after the growth (Fig. 3a).

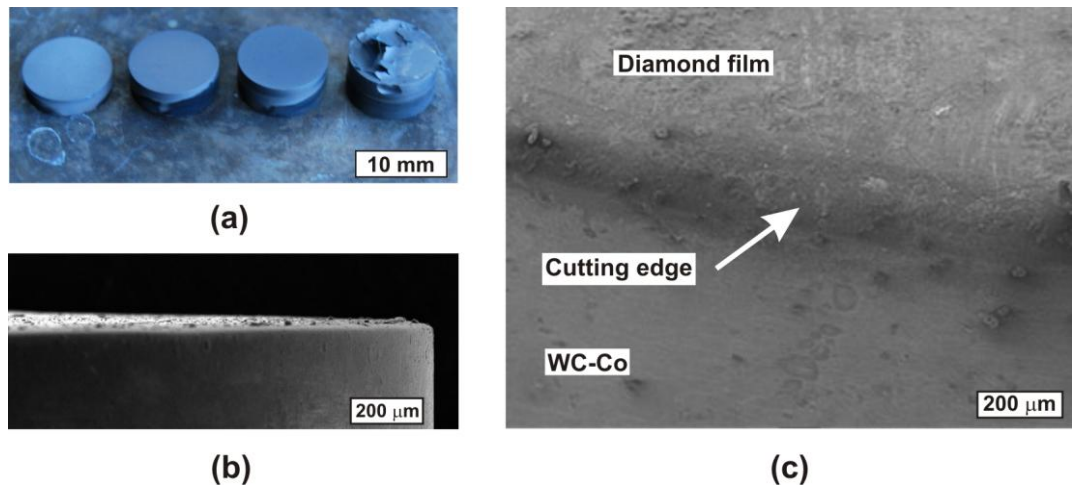

Fig. 3. Photo of WC-Co/W/diamond samples with height h of 2.5, 3.5, 4.5, 5.5 mm – second CVD series (a); SEM photos of the edge of sample with h = 4.5 mm (b) at grazing angle of 10 degrees and at angle of about 45 degrees (c) – for sample of the third CVD series.

Perhaps, delamination is caused by temperature rise, which results in (i) an increase in thermal stress due to the difference in the thermal expansion coefficients of diamond ($10^{-6}$ $K^{-1}$) and tungsten carbide–cobalt hard alloy ($4.8 \cdot 10^{-6}$ $K^{-1}$) and (ii) an increase in diffusion of cobalt through the tungsten barrier layer. The distance range Δh is likely to be wider; however, the determination of its boundaries requires additional experiments. Obviously, one can choose an optimal temperature by changing the distance between the substrate and plateholder surfaces during MWCVD growth of diamond. This will allow effective control of the growth rate of a uniform or a gradient structure of polycrystalline diamond films.

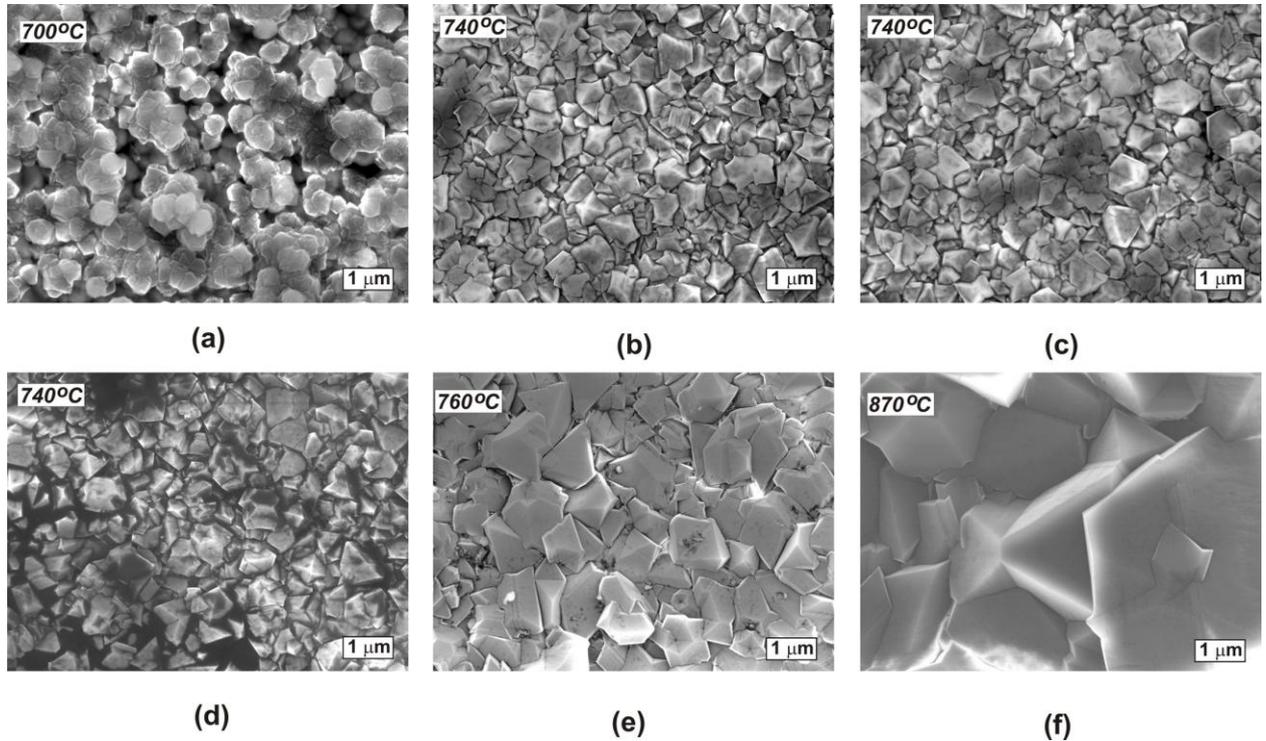

Fig. 4. SEM images of diamond films deposited at different temperatures and distances between the substrate and plateholder surfaces: Δh = − 1 mm, T(side) = 700°C, half of the radius (a); Δh = 0 mm, T(side) = 740°C, center (b); Δh = 0 mm, T(side) = 740°C, half of the radius (c); Δh = 0 mm, T(side) = 740°C, periphery (d); Δh = 1 mm, T(side) = 760°C, half of the radius (e), Δh = 2 mm, T(side) = 870°C, half of the radius (f). Micrographs (a), (e), (f) – second CVD series, and micrographs (b), (c), (d) – third CVD series.

All the films are nano- or microcrystalline with octahedral facets, typical of gray diamond, formed mainly by {111} planes with inherent twinning [19]. The mean grain size $d_{grain}$ increases with temperature (Table 1). We calculated $d_{grain}$ using SEM images (Fig. 4) at half the distance between the center and the edge of substrate. On the lowest sample (h/d = 0.25, T(side) = 700°C), the diamond film didn't become continuous and replicates the shape of micrograins of WC-6% Co hard alloy after selective etching in Murakami and Caro reagents. On other samples, the film surface was uniform; however, on the highest sample (Δh = 2 mm) deposited at 870°C, the film partially delaminated due to overheating. In the optimum growth region (T(side) = 740-760°C), we obtained a fairly uniform grain size: standard deviation $\sigma_{n-1}$ = 11% (Δh = 0 mm) and $\sigma_{n-1}$ = 12% (Δh = 1 mm). The results of optical profilometry are demonstrated in Table 2.

We couldn't measure roughness of the sample with Δh = 2 mm due to delamination of the diamond film after cooling from T(side) = 870°C down to room temperature and handling with it. It is interesting that the least mean roughness $R_a$ (Table 2) is observed at growth temperature of T(side) = 740°C, when sample and plateholder heights are the same (thickness of 0.6 μm, Δh = 0 mm, periphery): $R_a$ = 241 nm, rather than in the case of Δh = – 1 mm, when the growth temperature is the lowest (Table 1). At the same time, relations between roughness and grain size for different cases vary widely from $R_a = 0.33 \cdot d_{grain}$ (thickness of 4.1 μm, Δh = 1 mm, center) and $R_a = 0.44 \cdot d_{grain}$ (thickness of 1.2 μm, Δh = 0 mm, periphery) to $R_a = 1.68 \cdot d_{grain}$ (thickness of 0.6 μm, Δh = -1 mm, periphery). The latter case can be explained by the film discontinuity as at the periphery of the lowest (h = 2.5 mm) sample grain size was the smallest (240 nm) which indirectly testify to the lowest temperature near the sample edge. Increase in roughness leads to friction and accordingly wear increase [2, 20]. Consequently growth conditions for Δh = 0 mm possess an advantage over Δh = – 1 mm, 1 mm or 2 mm. The smallest grain size at the periphery of the highest sample (h = 5.5 mm) should be attributed to increased etching of diamond crystallites in hydrogen plasma in the nucleation and growth stages because of edge overheat which is seen at Fig. 1c. The relation between root-mean-square roughness $R_{rms}$ and $d_{grain}$ of so called black and white polycrystalline CVD diamond is found out to be respectively $R_{rms} = 1.1 \cdot d_{grain}$ and $R_{rms} = 0.25 \cdot d_{grain}$ [21]. The same relation has intermediate values in our case since gray diamond characteristics are intermediate between characteristics of black and white diamonds: $R_{rms} = 0.54 \cdot d_{grain}$ for Δh = 0 mm and $R_{rms} = 0.56 \cdot d_{grain}$ for Δh = 1 mm (Table 2).

Table 2. Roughness and grain size of diamond films grown in plateholder pockets on WC-6% Co substrates with different heights. Frame size for roughness measurements is 178×133 μm. Values for Δh = – 1 mm, 0 mm, 2 mm are from the second CVD series, and values for Δh = 1 mm are from the third CVD series.

| Distance between the substrate and plateholder surfaces Δh, mm | Roughness R(a), R(rms) and grain size d(grain) | | | | | | | | |
|---|---|---|---|---|---|---|---|---|---|
| | Sample center | | | Half of the radius | | | Sample periphery | | |
| | R(a), nm | R(rms), nm | d(grain), nm | R(a), nm | R(rms), nm | d(grain), nm | R(a), nm | R(rms), nm | d(grain), nm |
| -1 (h = 2.5) | 270 | 447 | 300 | 252 | 401 | 400 | 402 | 546 | 240 |
| 0 (h = 3.5) | 216 | 269 | 510 | 241 | 293 | 550 | 207 | 263 | 470 |
| +1 (h = 4.5) | 334 | 427 | 1020 | 506 | 584 | 890 | 469 | 600 | 1000 |
| +2 (h = 5.5) | – | – | – | – | – | 2170 | | | |

The space resolution of the pyrometer Micron M770 we used for temperature measurements from the top of the substrates (Fig. 1a,e) didn't enable to distinguish different points on the surface. However, we can indirectly compare the temperature there. Grain size increases with deposition temperature. The edge effect leads to grain rise on the substrate periphery where the temperature is higher. It turned out that there is no clear regularity of grain size at the surface in our case. The largest grains can be at the half of the radius (Δh = -1 mm, 0 mm, 2 mm) or at the center (Δh = +1 mm) in contrast with the deposition in open-type substrate holder [8], in which case the largest grains are located at the sample periphery. We don't see that usual dependence for the samples of Δh = -1, 0, 1 mm which correspond with absence of noticeable edge effect under visual observation (Fig. 1c).

Grain size of diamond films grown in open-type sample holder on WC-Co substrate is not uniform [8] because of the edge effect. Roughness and grain size (Table 2) of diamond films on a substrate with Δh of – 1 mm in closed-type plateholder is rather uniform and amounts to $R_a$ = 220 ± 20 nm (±9%) and $d_{grain}$ = 510 ± 40 nm (±8%). On the sample grown at Δh of 1 mm, the roughness and grain size on the surface increases respectively to $R_a$ = 500 ± 70 nm (±14%) and $d_{grain}$ = 950 ± 70 nm (±7%). These values are much less than for MPCVD on open-type holder [8] where diamond grain size varied widely from 2.3 μm at the center to 10.1 μm at the edge of a cutting insert i.e. $d_{grain}$ = 6.2±3.9 μm (±63%). Roughness and grain size of diamond films deposited at Δh = – 1 mm is not so uniform and reaches $R_a$ = 327 ± 80 nm (±23%) and $d_{grain}$ = 320 ± 80 nm (±25%). Such discontinuity outside the optimum growth region can be explained by growth rate owing to reduced deposition temperature. If the deposition time were longer, the discontinuity of the sample Δh = -1 mm would reduce. The obtained range Δh = 0…1 mm values meet the requirements of the ISO 13399 standard for various types of carbide inserts, both for general-purpose and special-purpose tools, namely 3-d pattern on lateral plane are not higher than 1 mm.

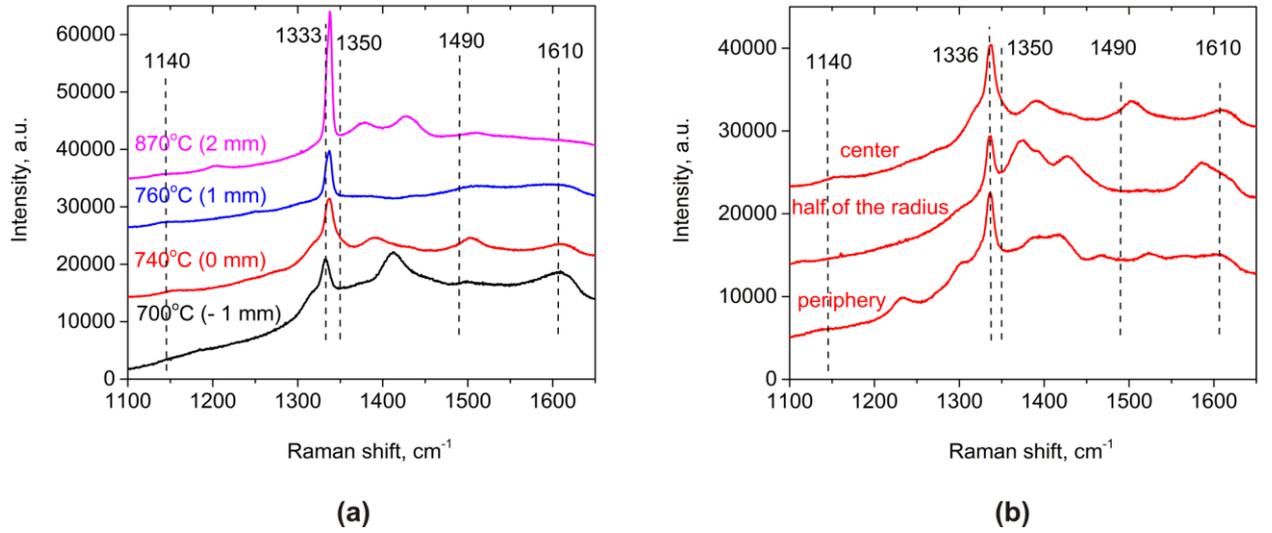

Fig. 5. Raman spectra of polycrystalline diamond films synthesized at temperatures T(side) from 700°C to 870°C and distances between the substrate and plateholder surfaces Δh from – 1 mm to +2 mm (a) and 3 spectra taken at different points on the sample grown at Δh = 0 mm and T(side) = 740°C (b). Spectra for Δh = – 1 mm, 0 mm, 2 mm are from the second CVD series, and spectrum for Δh = 1 mm is from the third CVD series.

Raman spectra of the films (Fig. 5a) deposited at temperatures of T(side) = 700, 740, and 760°C (samples with Δh = – 1 mm, 0 mm and 1 mm, respectively) contain transpolyacetylene (TPA) bands at 1490-1500 cm$^{-1}$ (700°C) and 1140 cm$^{-1}$ (740 and 760°C). These TPA bands are characteristics of nanocrystalline diamond with large content of sp$^2$ phase. The diamond Raman line at 1333 cm$^{-1}$ increases with deposition temperature as usual (Fig. 5a). Its full width at half maximum (FWHM) decreases from 12.7 cm$^{-1}$ (Δh = – 1 mm) to 5.6 cm$^{-1}$ (Δh = 2 mm). The intensity of the graphite G band decreases with temperature, which reflects the growth of the grain size and increase of sp$^3$ phase content in the samples. The D band is seen only in the first sample (at 700°C), and the G band (1580-1610 cm$^{-1}$) in the fourth sample (at 870°C) disappears completely. There are some indefinite lines at 1370-1430 cm$^{-1}$ (Fig. 5), probably belonging to amorphous carbon typical for WC-Co/diamond samples [20]. According to [24], relative intensity for diamond to graphite bonded carbon with 473 nm excitation is about 1:75. Appling this relation, we can deduce that real sp$^3$ phase proportion rises with temperature from 98.5% at. (Δh = – 1 mm) to 99.7% at. (Δh = 2 mm).

All samples exhibit sharp diamond peaks from frequency $v_x$ = 1333.2 cm$^{-1}$ at T(side) = 700°C to $v_x$ = 1337.3 cm$^{-1}$ at T(side) = 870°C. The peak for non-stressed diamond ($v_{ns}$) is at 1332 cm$^{-1}$. This blue shift of diamond line reflects high compressive stress inside the films caused by the difference in the thermal expansion of WC-6% Co ceramic and diamond. The stress can be estimated [22, 23] by the formula $\sigma$ [GPa] $= -0.567(\nu_{ns} - \nu_x)$ [cm$^{-1}$]. In our samples, it varies from $\sigma = 0.68$ GPa for T(side) = 700°C to $\sigma = 2.2, 2.5$ and 3.0 GPa for T(side) = 740°C, 760°C, or 870°C, respectively. The higher growth temperature, the higher $\nu_x$ shift and compressive stress of CVD diamond film. We can indirectly estimate growth temperature, for example on the diamond surface grown at $\Delta h = 0$ mm (Fig. 5b). The $\nu_x - \nu_{ns}$ shift on the center/half of the radius/periphery amounts accordingly to 4.3/3.5/3.9 cm$^{-1}$ indicating the absence of overheat at the edge (Fig. 5b). Therefore, Raman shift allows us to conclude: T(half of the radius) < T(edge) < T(center) for $\Delta h = 0$ mm sample. Both stress (Raman $\nu_x$ shift) and grain size or roughness values of this sample testify to mean or the least edge heating, respectively.

There are two factors leading to delamination of the diamond film. The first is thermal expansion mismatch meaning that tungsten carbide compresses up to five times faster than grown CVD diamond upon it when cooling down to room temperature. The second factor is catalysis of graphitization by cobalt that dissolves actively carbon atoms [2] of growing thermodynamically unstable diamond under CVD temperatures (700-900°C). Carbon solubility in cobalt decreases with temperature decrease. After MPCVD, Co releases excess C atoms under cooling of WC-Co substrate with polycrystalline diamond which generates thermodynamically stable graphite interlayer between tungsten carbide and diamond. This interlayer impairs adhesion and, together with fivefold difference in compression rate of WC-Co and diamond provides delamination.

There is no Co impurities anywhere on the diamond film surface of samples $\Delta h = -1$ mm and 0 mm, therefore we succeed in Co blocking. Cobalt appears on the diamond film surface of samples $\Delta h = 1$ mm and 2 mm. XRF detected 0.19% at. of Co at the edge of the sample $\Delta h = 1$ mm, 0.02% at. of Co at the center, and no Co impurities at the half of the radius. The higher WC-Co/W/diamond temperature, the easier Co diffusion towards diamond surface. This is the third way to estimate deposition temperature. We conclude therefore that T(half of the radius) < T(center) < T(edge) for $\Delta h = 1$ mm (XRF estimation). From $d_{grain}$ estimation (Table 2) we deduce another result: T(half of the radius) < T(edge) < T(center). In both cases, temperature at the half of the radius is appears to be minimal which agrees with result for $\Delta h = 2$ mm sample where diamond film retained only at the half of the radius which indicates there minimal graphitization caused by cobalt diffusion. XRF detected 0.04% at. of Co at the half of the radius on the diamond surface ($\Delta h = 2$ mm), 0.32% at. of Co on the center (on the WC-Co/W surface, the diamond film delaminated there) and 1.73% at. on the periphery (WC-Co surface, the film delaminated with W layer) (Fig. 3a). Accordingly to XRF estimation for $\Delta h = 2$ mm sample, T(half of the radius) < T(center) < T(periphery) which testifies about the edge effect. This observation we can't compare with $d_{grain}$ or Raman shift estimations because of the film delamination. Thus, all these three ways of temperature estimation (roughness or grain size, stress, Co concentration) argue for the films uniformity and absence of the edge overheating within range of distances Δh = – 1…1 mm.

Wear resistance and cutting efficiency of high aspect ratio substrates with (by an example of the sample with height of 4.5 mm, Δh = 1 mm, T(side) = 760°C, from the third CVD series) and without CVD diamond coating was measured by machining A390 silumin containing 18% wt. of silicon carbide.

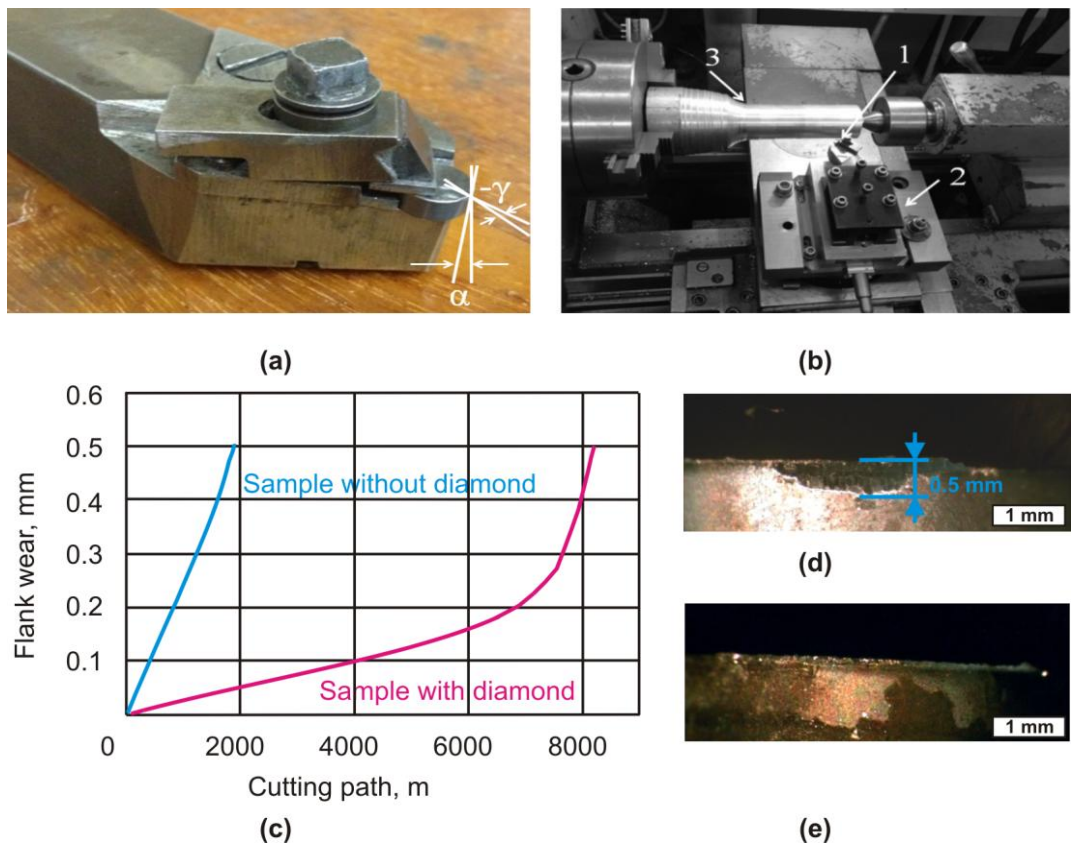

Fig. 6. Experiment on cutting A390 alloy by WC-6% Co plates with CVD diamond coating: photo of a test sample with orientation angles γ and α in the holder (a); photo of a test facility for cutting A390 silumin: (1) cutting plate, (2) dynamometer, and (3) work piece (b); wear curves of two WC-6% Co plates (h = 4.5 mm) with and without diamond coating (c); wear of edge for cutting path 1890 m: photo of WC-6% Co plate (h = 4.5 mm) without (d) and with (e) diamond coating. Pictures (c) and (e) are for the sample of the third CVD series.

The measured curvature radius, which characterizes the sharpness of the cutting edge of a plate without coating was $\rho$ = 50 μm and that of a plate with diamond coating (Δh = 1 mm) was $\rho$ = 30 µm; thus, the deposition of diamond increased the sharpening of the cutting edge by a factor of 1.6, as measured with a "Calibre 252" profilometer. It can be explained by peculiarity of the diamond film deposition on the cylinder edge. There is a "cap" of the film on upper 200 microns (Fig. 3c) of the cylinder wall. Since the original sample had the shape of a cylinder, rather than of a cutter, we manufactured a special holder to carry out the tests that allowed one to orient the cutting edge at angles of $\alpha = 6°$ and $\gamma = -6°$ (Fig. 6a). The sample machined on a "1K62" automatic screw machine (Fig 6b). The cutting force (Table 3) measured on the diamond coated plate $F_{cut}$ = 142 N was 52% of the cutting force of the non-coated plate. This allowed us to increase the cutting path length by a factor of 4.3: from 1890 to 8150 m (Fig. 6c). The flank wear after cutting path of 1890 m (Fig. 6d) reaches the limit (0.5 mm) for the uncoated sample while it is negligible (Fig. 6e) for the diamond coated one. Cutting force is relatively high but we measure it for unoptimized cutter shape. In the case of milling tests cutting force is lower by 1-2 orders of magnitude [2]. $F_{cut} > 100$ N usually results in diamond coating delamination when milling, but good adhesion allows the film to sustain cutting.

Table 3. Wear resistance and cutting force of WC-6% Co plates (h = 4.5 mm) with and without diamond coating: cutting of A390 alloy.

| Characteristic | WC-6% Co plate without coating | WC-6% Co plate with diamond coating |
|---|---|---|
| Wear area, mm$^2$ | 0.45 | 0.4 |
| Path for the limit wear of 0.5 mm, km | 1.89 | 8.15 |
| Volume of removed shavings, cm$^3$ | 142.5 | 611 |
| Specific wear, mm$^2$/km | 0.237 | 0.049 |
| Amount of wear, mm$^3$ | 0.0106 | 0.0084 |
| Wear rate, mm$^3$/m$^3$ of shavings | 137 | 74 |
| Cutting force, N | 273 | 142 |

Flank wear decreased (Table 3) from 137 mm$^3$ of the initial WC-Co sample material per 1 m$^3$ of A390 alloy to 74 mm$^3$/m$^3$ for the diamond coated WC-Co sample. This value will increase when (i) growing thicker diamond film (4 µm is many times smaller amount than usually [20]) and (ii) using real inserts engineered for cutting.

**Conclusions**

Using a pocket-type substrate-holder (plateholder), we have demonstrated the elimination of the edge effect, which is typical of growth in an open substrate-holder, during the synthesis of diamond films on WC-6%Co plates with high aspect ratio (0.45 < h/d < 0.35), which provided uniform heating of substrates without overheating of their edges. We have found that the temperature, diamond growth rate, and the grain size in a polycrystalline film strongly depend on the distance Δh between plateholder and substrate surfaces. The temperature range that provides uniform deposition of a diamond film on WC-6%Co cutters in our CVD system is 740-760$^o$C for power of 2.9 kW and distances 0 ≤ Δh > 1 mm; in this case, the variation of roughness $R_a$ of the diamond film from the centre to the periphery does not exceed 15-16%. We have demonstrated that a variation of the distance Δh between substrate and plateholder provides an efficient method for controlling the temperature of the growth surface and the structure of polycrystalline diamond films.

The cutting force measured when machining by a WC-6%Co diamond-coated plate decreases twice as compared with that of the uncoated plate. This allowed one to enhance the cutting efficiency by a factor of 4.3 and increase the cutting path from 1890 m to 8150 m. The cutting force and the flank wear were decreased twice.

**Acknowledgements**

The authors are grateful to E. Zavedeev for the optical profilometry analysis, A. Rudenko for the SEM analysis, and to V. Yurov for assistance in some pictures preparation. This work was supported by Russian Science Foundation, grant No. 15-19-00279.